# The nature of the long time decay at a second order transition point

Moshe Schwartz
School of Physics and Astronomy
Tel Aviv University
Tel Aviv, Ramat Aviv, Israel

And

S. F. Edwards
Cavendish Laboratory
Maddingley Rd.
CB3 OHE, Cambridge, UK

We show that at a second order phase transition, of $\phi^4$ like system, a necessary condition for stretched exponential decay of the time dependent structure factor is obeyed. Using the ideas presented in this proof a crude estimate of the decay of the structure factor is obtained and shown to yield stretched exponential decay under very reasonable conditions.

The decay of disturbances in some physical systems is known to be a stretched exponential in time. The canonical list of those systems includes glasses, [1-3] polymer solutions [4,5] and such a form of decay is also characteristic of dielectric and viscoelastic relaxation. [6,7] In recent work [8,9] we have shown that this form of decay is much wider spread than expected. We claim, in fact, that such a decay is present in many non-linear systems such as magnets at their critical point, systems of the KPZ type etc. These systems share the property that a disturbance of wave vector $\vec{q}$, decays at equilibrium with a characteristic decay rate, $\omega_q$, that is proportional to $q^\mu$ with $\mu > 1$. Since, the problems involved are very difficult, the derivation has to rely on approximations. Therefore, it is very important to have some exact results that may also suggest an intuitive understanding of the stretched exponential decay. We start by giving an exact proof that the condition for a slow form of decay is fulfilled at the critical point of a system of the $\phi^4$ type. We continue then, by employing the same basic ideas, to argue that indeed the decay is a stretched exponential in time. The systems we consider here are very different than those considered so far in the context of stretched exponential decay and we are interested in $\vec{q}$ dependent correlation functions, yet, the final step in the derivation is similar in its mathematical structure to other derivations in which stretched exponential decay has been obtained. [10-13]

The system we have in mind is a classical $\phi^4$ system at its critical point or any generalization of it. We will also assume that the classical Hamiltonian, $W$, that gouverns the statics is even in the field $\phi$. It is well known that Langevin dynamics of the system leads to a Schroedinger like evolution equation [14-17]

$$\frac{\partial \psi}{\partial t} = -H\psi, \qquad (1)$$

where $\psi$ is related to the probability distribution of obtaining a given field configuration $\{\phi_q\}$ at time $t$, $P\{\phi_q,t\}$, by $P\{\phi_q,t\} = \exp[-W/2]\psi\{\phi_q,t\}$ and where the "Hamiltonian"

$$H = -\sum \lambda_q \{\frac{\partial^2}{\partial \phi_q \partial \phi_{-q}} + \frac{1}{2}\frac{\partial^2 W}{\partial \phi_q \partial \phi_{-q}} - \frac{1}{4}\frac{\partial W}{\partial \phi_q}\frac{\partial W}{\partial \phi_{-q}}\} \qquad (\lambda_q = \lambda_{-q} > 0) \qquad (2)$$

is Hermitian, non negative and its only eigenfunction with eigenvalue zero, is the ground state $\psi_G \equiv P_{eq}^{1/2} \propto \exp[-W/2]$. In the following we will chose $\lambda_q$ to be proportional to $q^\theta$ with $\theta$ that is not too negative. (The meaning of the last phrase will become clear later). Past studies of the dynamics of the $\phi^4$ system concentrated on $\theta = 0$ and $\theta = 2$. [18-19]

The persistence of a disturbance of wave vector $\vec{q}$ at equilibrium is described by the normalized time dependent structure factor

$$\alpha_q(t) = \frac{\Phi_q(t)}{\Phi_q(0)} = \frac{<\phi_{-q}(0)\phi_q(t)>}{<\phi_{-q}(0)\phi_q(0)>} \ ,$$

where the averages on the right hand side of the above mean that $\phi_{-q}$ is measured in equilibrium at time $t=0$ then the system is allowed to evolve freely and $\phi_q$ is measured at time $t$. Next, we express $\alpha_q(t)$ in terms of the eigenfunctions and eigenvalues of the "Hamiltonian" $H$. Since $H$ commutes with the momentum operator, the eigenfunctions of $H$ can be chosen also to be eigenfunctions of the momentum operator. Let $\{\psi_{\vec{q},\beta}\}$ be the set of normalized eigenfucntions of $H$, that are also eigenfunctions of the momentum operator with eigenvalue $\vec{q}$. The corresponding set of eigenvalues of $H$ is $\{\lambda_{q\beta}\}$. We find that

$$\alpha_q(t) = \frac{\sum_\beta |(\psi_{\vec{q},\beta},\phi_q\psi_G)|^2 \, e^{-\lambda_{q\beta} t}}{\sum_\beta |(\psi_{\vec{q},\beta},\phi_q\psi_G)|^2} \tag{3}$$

It is obvious that if the smallest eigenvalue in the set $\{\lambda_{q\beta}\}$, $\lambda_{qo}$, is really positive, $\alpha_q(t)$ cannot decay slower than exponential. Our first goal is to show that to zero order in $\Omega$, the volume of the system, $\lambda_{qo}$ is indeed zero at the critical point.

Consider the set of unnormalized eigenfunctions of the momentum operator with momentum $\vec{q}$,

$$\chi(\vec{q};\vec{\ell}_1,...,\vec{\ell}_n) = \Pi \phi_{\vec{\ell}_i} |\psi_G> \qquad (\sum \vec{\ell}_i = \vec{q}) \tag{4}$$

(To simplify we restrict ourselves in the following to sets $\{\vec{\ell}_i\}$ that do not have a subset that sums up to zero. This restriction can be avoided in a more detailed presentation). The expectation value of $H$,

$$\mu^{(n)}(\vec{q},\vec{\ell}_1,...,\vec{\ell}_n) = \frac{<\chi(\vec{q};\vec{\ell}_1,...,\vec{\ell}_n)|H|\chi(\vec{q};\vec{\ell}_1,...,\vec{\ell}_n)>}{<\chi(\vec{q},\vec{\ell}_1,...,\vec{\ell}_n)|\chi(\vec{q},\vec{\ell}_1,...,\vec{\ell}_n)>} \ , \tag{5}$$

is an exact upper bound on $\lambda_{qo}$ by the variational theorem (applied to the sector of states with momentum $\vec{q}$). The expectation value $\mu$ can be easily calculated, using the identity

$$<\psi_G | \Lambda_{-q} H \Lambda_q | \psi_G> = \frac{1}{2} <\psi_G | [\Lambda_{-q},[H,\Lambda_q]] | \psi_G> \qquad (6)$$

The result is

$$\mu^{(n)}(\vec{q};\vec{\ell}_1,...,\vec{\ell}_n) = \sum_{i=1}^{n} \lambda_{\ell_i} / \Phi_{\ell_i}(0) + O(1/\Omega) , \qquad (7)$$

so that in the limit of the infinite system the sum on the right hand of the above yields the exact result. Note, that although the final result may seem to be the result of a Gaussian approximation it is in fact the full exact result. The non Gaussian corrections vanish when the volume of the system tends to infinity. At the transition $\Phi_\ell(0)$ is proportional to $\ell^{-2+\eta}$, so that the term that is zero order in $\Omega$ on the right hand side of eq.(7) has the form $\sum |\ell_i|^\Gamma$ ($\sum \vec{\ell}_i = \vec{q}$). It is straightforward to show that if $\Gamma \geq 1$ such a sum attains its minimal value when $\vec{\ell}_i = \vec{q}/n$. The value of the sum at the minimum is $\mu^{(n)}(q) \alpha \ n^{-(1+\theta)+\eta} q^{2+\theta-\eta}$. If $\theta$ is not too negative $1+\theta-\eta$ is positive and $\mu^{(n)}(q)$ tends to zero as $n$ tends to infinity, for all $\vec{q}$. This implies at once that $\lambda_{qo}$ is indeed zero. What we have shown is that the (positive) spectrum $\{\lambda_{q\beta}\}$ is not bound from below by any positive value. We have thus proven that the necessary condition for the existence of a stretched exponential decay holds at the point of transition.

We turn now to obtaining a crude estimate of $\alpha_q(t)$ at long times, that is based on the simple ideas that have already been presented above. We will assume that the system has excitations carrying momentum $\vec{\ell}$ and energy $\omega_\ell = B\ell^z$ and that all the states of the system can be approximately described as a collection of such excitations, so that each state can be tagged by the total number of excitations present and the "energy" of a state carrying total momentum $\vec{q}$ is given approximately by

$$\lambda^{(n)}(\vec{q};\vec{\ell}_i,...,\vec{\ell}_n) = \sum_{i=1}^{n} \omega_{\ell i} \qquad (8)$$

This assumption is motivated by the form of the trial functions (eq. 5) and the corresponding diagonal elements of the Hamiltonian $H$, (eq. 7).

The decay function $\alpha_q(t)$ is given by

$$\alpha_q(t) = \sum_{n=1,3}^{\infty} \int \Pi d\vec{\ell}_i \ \delta(\sum \vec{\ell}_i - \vec{q}) |a_n(\vec{\ell}_1,...,\vec{\ell}_n)|^2 e^{-\sum \omega_{\ell i} t} \qquad (9)$$

By the variational theorem it is obvious that $z > 1$, so that the minimal value of $\lambda^{(n)}(\vec{q}; \vec{\ell}_i, ..., \vec{\ell}_n)$ is $Bn^{1-z}q^z$ and this value is attained at the point $\{\vec{\ell}_i = \vec{q}/n\}$.
Consider the quantity

$$A_n(q,t) = \int \prod_{i=1}^{n} d\vec{\ell}_i \delta(\sum \vec{\ell}_i - \vec{q}) |a_n(\vec{\ell}_1, ..., \vec{\ell}_n)|^2 e^{-\sum \omega_{\ell_i} t} \qquad (10)$$

It is clear that the most dominant contribution to $A_n(q,t)$ comes from the vicinity of the point $\{\vec{\ell}_i = \vec{q}/n\}$. Standard asymptotics yields the large $t$ behaviour of $A_n(q,t)$

$$A_n(q,t) = \exp[-f(n) - \frac{d}{2}(n-1)\ell n(\omega_q t) - n^{1-z}\omega_q t] \quad . \qquad (11)$$

In obtaining the above we also required that $A_n(q,t) = A_n(\omega_q t)$. The long time behaviour of $\alpha_q(t)$ is now determined by obtaining the most dominant $A_n(\omega_q t)$ from the equation

$$-f'(n) - \frac{d}{2}\ell n(\omega_q t) + (z-1)n^{-z}\omega_q t = 0 \qquad (12)$$

Estimation of the function $f(n)$ is quite complicated and beyond the scope of this paper. It will turn out, however, that much can be said about the decay function without knowing the detailed form of $f(n)$. Assume first that $f(n)$ is either linear in $n$ for large $n$ or that $\lim_{n \to \infty} \frac{f(n)}{n} = 0$. If this is the case the first term on the left hand side of eq. (12) can be ignored and the conclusion is that the dominant contribution comes from $n = n^*$, where

$$n^* = [\frac{2(z-1)}{d} \frac{\omega_q t}{\ell n \omega_q t}]^{\frac{1}{z}} , \qquad (13)$$

So that the decay function $\alpha_q(t)$ can be written as $\exp[-F(\omega_q t)]$, where to leading order

$$F[\omega_q t] \propto [\omega_q t]^{\frac{1}{z}} [\ell n \omega_q t]^{(1-\frac{1}{z})} . \qquad (14)$$

Apart from the logarithmic correction the stretched exponential decay obtained here is exactly the form we found by using entirely different methods, [8,9].

Now it is clear that the decay becomes faster when $f(n)$ is increased. Therefore, it is not possible that the two point function decays slower than the decay given by eq. (14).

Consider next what happens if $f(n) \propto n^\alpha$ with $\alpha > 1$. It is easily verified that in such a case the dominant terms on the left hand side of eq.(12) are the first and the last and $n^*$ giving the main contribution to the sum is

$$n^* = [\frac{z-1}{\alpha} \omega_q t]^{\frac{1}{\alpha+z-1}} , \qquad (15)$$

and

$$F(\omega_q t) \propto [\omega_q t]^{\frac{\alpha}{\alpha+z-1}} . \qquad (16)$$

We see that even here the final result is a stretched exponential decay. The stretched exponent is larger than before but it is still stretched. Only in the limit $\alpha \to \infty$ we recover the unstretched exponential. It is interesting to consider the case where indeed the decay is exponential ($\alpha = \infty$) to understand why it cannot be generic. Consider the "Hamiltonian" in (eq.2) for the Gaussian model (no $\phi^4$ term) at its transition point. The "excitation energy" $\omega_q$ is given by $\omega_q = vq^2 \lambda_q$ and the decay function is $\alpha_q(t) = \exp[-\omega_q t]$. The reason is that $\phi_q |0>$ is an exact eigenstate of the "Hamiltonian". Consequently, the sum in eq. (9) over states with $n$ excitations is limited to $n=1$. This sharp cut-off can be interpreted as $\alpha = \infty$. In the non linear case $\phi_q |0>$ is far from being an exact eigenstate of the "Hamiltonian" so that it is extremely unlikely that $\alpha = \infty$. Furthermore, even if a non-linear model could be constructed to yield $\alpha = \infty$, this property would not be stable against small changes in the model. Namely $\alpha \neq \infty$ is the generic case for non-linear systems. In fact, we have good reasons to believe that the generic situation is characterized not only by $\alpha \neq \infty$ but by $\alpha \leq 1$, so that eq. (14) describes the decay correctly.

We hope to present a detailed estimation of $f(n)$ in the near future.